\documentclass{article}

\usepackage[utf8]{inputenc}
\usepackage[british]{babel}
\usepackage{CJKutf8}

\usepackage{microtype}

\usepackage{tikz}
\usepackage{graphicx}
\usepackage{environ}
\usepackage{pdfpages}
\usepackage{mdwlist}
\NewEnviron{scale}[1]{%
  \begin{tikzpicture}
    \node[scale=#1]{%
      \BODY
    };
  \end{tikzpicture}
}

\usepackage[hmarginratio=1:1,top=32mm,columnsep=20pt]{geometry}
\usepackage{multicol}
\usepackage[hang, small,labelfont=bf,up,textfont=it,up]{caption}
\usepackage{booktabs}
\usepackage{float}
\usepackage[colorlinks,citecolor=blue]{hyperref}
\usepackage{listings}
\usepackage{amsmath}

\usepackage{abstract}
\usepackage{titlesec}
\titleformat{\section}[block]{\large\scshape\centering}{\thesection.}{1em}{}
\titleformat{\subsection}[block]{\large}{\thesubsection.}{1em}{}

\title{\vspace{-15mm}\fontsize{16pt}{10pt}\selectfont
\textbf{Extending the Learning by Teaching Canvas System: Maximising Academic 
Learning Time}}

\author{\begin{tabular}{c c c}
    Alexander Berntsen \\
    \href{mailto:alexander@plaimi.net}{alexander@plaimi.net}
\end{tabular}
}

\graphicspath{{fig/}}

\begin{document}
\maketitle
\begin{abstract}
\noindent Time plays an integral role in the realm of e-learning modules. plaimi 
previously described a canvas for composing e-learning modules. By 
transitivity, time needs to play an integral role in this canvas. This paper 
investigates some of the ways it can play said role. Four angles are 
considered: insights offered by viewing a module composition as a chronicle of 
modules, the importance of length estimation in time allocation, heightening 
retention via spaced repetition, and synchronisation attempts at facilitating 
collaborative learning. The features discovered by this investigation are 
discussed in a principled learning context, which particularly emphasises 
academic learning time. Some concrete suggestions are made; to implement: 
order-awareness, user estimation of module length, 
spaced-repetition-awareness, and post-module self-assessment. Suggestions for 
further research are also given.

\end{abstract}
\tableofcontents
\listoffigures
\newpage
\begin{multicols}{2}
    \section{Introduction}
In an effort to foster learning by teaching, plaimi have previously described 
a canvas for intuitively composing e-learning 
modules\cite{berntsen2015enabling}. This system indirectly emphasises 
chronology, which plaimi previously explored in the tempuhs 
system\cite{berntsen2014tempuhs}.

The canvas system lets users drag and drop e-learning modules onto it, and 
then arrange the data flow of the system, thereby effectively arranging the 
modules into a chronology of modules (a composition of modules). By looking at 
the modules as a chronology, and considering the role time plays in principled 
learning, there are several insights available to us.

This paper describes some such insights. It motivates the insights, and 
explore them in some detail. This includes elaborating and elucidating the 
concepts, as well as giving some notes on their potential implementation. 
There are numerous challenges that the papers elects to not ignore, and seeks 
to mitigate.

Since we are designing a system for learning, it is important that any 
features we consider for inclusion have a sound scientific foundation. The 
insights offered and features discussed are thus considered in a scientific 
context.

There are four angles explored by the paper. Insights afforded by chronicling 
and ordering are presented in Section~\ref{chronology}. Estimation, both by 
way of users estimating their modules' time frame, and by the system 
automatically estimating it, is discussed in Section~\ref{estimation}. Spaced 
repetition as a way of improving retention is explored in 
Section~\ref{repetition}. Finally, features related to synchronous 
collaboration and timeslot synchronisation are described in 
Section~\ref{synchronisation}. We also take the time to make a few remarks 
regarding future research in Section~\ref{further}.

\section{Motivation}
When designing a learning experience, it is essential to consider time. This 
includes an awareness of allocated, engaged, and academic learning time (ALT), 
lest dead time incurs for want of understanding. Allocated time is the amount 
of time allocated for learning. Engaged time is time spent actively attempting 
to learn. ALT is time spent engaged in appropriate learning that leads to high 
levels of success\cite{cotton1990educational}; i.e.\ time spent in the flow. 
Flow is the state of being fully immersed and focused on an 
activity\cite{murphy2011games}.

There is a very slight but persistent correlation between allocated time and 
achievement\cite{cotton1981time, walberg1988synthesis, cotton1990educational}. 
Engaged time is modestly correlated with achievement\cite{cotton1981time, 
sanford1983time, cotton1990educational}. ALT leads to more 
learning\cite{walberg1988synthesis}, and the rate of ALT is \emph{highly} 
correlated with achievement\cite{cotton1981time, sanford1983time, 
walberg1988synthesis, cotton1990educational}. Related to our canvas system we 
also take note that interactive engaged time lead to higher achievement than 
non-interactive engaged time\cite{sanford1983time, cotton1990educational}.

Pre-laptop era research demonstrates unequivocally that school pupils only 
spend roughly half of their in-class time engaged in 
learning\cite{cotton1990educational}. Laptop era research shows that students 
with laptops, compared to those without, spend more time engaged in learning, 
develop better critical thinking skills, and are more self-reliant. 
Additionally, laptop users are significantly higher-achieving than their 
non-laptop-using counterparts\cite{cengiz2005learning}. There is no indication 
that this research should not extend to today's era of laptops coexisting with 
tablets and sophisticated mobile phones in the classroom.

When considering the canvas system, we need to seek not only to maximise 
engaged time, but also to allow e-learning modules authors to think about how 
their module will fit into allocated time. Furthermore, situated learning 
environment professionals, e.g.\ teachers at primary schools, necessarily need 
consider allocated classroom time when choosing which modules to use. 

Moreover, and more importantly, the system must strive for the maximisation of 
ALT by eliminating material that is either too easy or too difficult, or 
otherwise unsuitable to the learner. The system admits the possibility of 
non-linear adaptive compositions of e-learning modules --- and it would not be 
unfaithful to the original concept to attach the utmost importance to this 
goal --- leading to a great opportunity to further elevate the present ALT 
ratio both in and out of classrooms.

In addition to maximising learning, we also seek to maximise motivation 
(willingness to engage) and minimise procrastination (unwillingness to engage; 
the absence of (self-regulated) performance\cite{lee2005relationship}). First, 
we suppose that our canvas system is a gamified system, and thus it is 
inherently comparable to a game in several ways\cite{deterding2011game}. This 
is intended design\cite{berntsen2015enabling}. Furthermore, designing 
e-learning modules is a subset of instructional design, which is fundamentally 
similar to game design\cite{murphy2011games}. Then we accept that ALT in our 
system is a form of flow. This is a reasonable conjecture, because flow state 
works by precisely the same mechanics as ALT: performing at the edge of one's 
competency, guided by feedback\cite{rutledgepositive}. This is altogether the 
point of ALT, as designed and desired to manifest in our gamified system. From 
this follows several insights.

Flow has a number of different desirable properties. It is intrinsically 
linked to motivation and widely accepted as one of the fundamental reasons 
people play games\cite{murphy2011games}, and thus an emphasis on flow might 
cause people to use our canvas system. It follows immediately from the law of 
readiness (and indirectly from the law of effect) that learners learn best 
when motivated\cite{murphy2011games}, and the whole point of our system is for 
our users to learn. Additionally, presence of flow is significantly negatively 
correlated with procrastination, and absence of flow is significantly 
positively correlated with procrastination\cite{lee2005relationship}.

Consequently, we must conclude that ALT --- and by extension time --- as a 
concept is intrinsic to our system.

\section{Ideas}
\subsection{Chronicling}
\label{chronology}
\begin{figure}[H]
\begin{centering}
\begin{tikzpicture}[node distance = 6 em, auto]
\node(m){m};
\node[right of = m](n){n};
\node[right of = n](i){};
\node[above of = i](o){o};
\node[below of = i](p){p};
\node[right of = i](q){q};

\draw[thick,->](m)--(n);
\draw[thick,->](n)--(o);
\draw[thick,->](n)--(p);
\draw[thick,->](o)--(p);
\draw[thick,->](p)--(q);
\end{tikzpicture}
\caption{An imagined popular canvas}
\label{canvas}
\end{centering}
\end{figure}
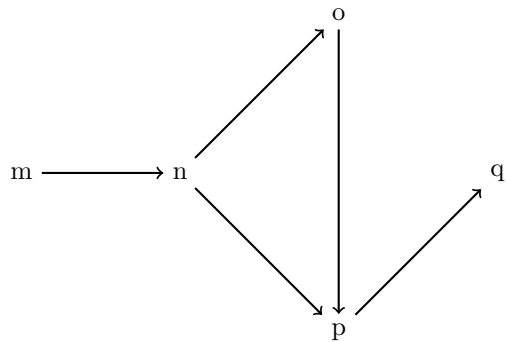

The canvas system may be aware of the chronology in a composition. After all, 
it is already there indirectly. As a motivation, let's consider an imagined 
popular canvas. Let $m$, $n$, $o$, $p$, and $q$ be modules. Let $m \to n$, $n 
\to o$, $n \to p$, $o \to p$, and $p \to q$ be possible flows, where $\to$ is 
a binary operator signifying that the user is sent from the module given as 
its left-hand side argument (lhs) to the module given as its right-hand side 
argument (rhs). This canvas is shown in Figure~\ref{canvas}.

Some self-evident properties here include that the successor of $m = n$, and 
the successor of $n = o \vert p$. Module relations are transitive with respect 
to order, so the successor of $n = q$ via the successor of $o = q$, and the 
successor of $p = q$ --- all the way down to the successor of $m = q$. 
Alternatively, by symmetry, we can say that the predecessor of $q = m$.

Ordering is trivial to store in module metadata, and via very simple 
mathematical operations we are afforded a lot of useful insights. Just from 
the above, several concrete features are easily imagined.

\begin{itemize*}
  \item We can suggest that authors of compositions that contain e.g.\ module 
$m$ might be interested in modules $n$, $o$, $p$, and $q$.
   \item Authors with modules $m$ and $q$ on the canvas could be recommended 
   to insert modules $n$ and $o$ in-between them. This is particularly 
   interesting if the system has other useful metadata. E.g.\ if $q$ is 
   generally considered to be very difficult, and $n$ and $o$ is considered to 
   augment $m$ significantly in preparing for $q$, $n$ and $o$ should be 
   strongly recommended to the author of a canvas with $m$ and $q$ in them.
   \item Authors with modules $m$, $n$, $o$, and $p$, on their canvas could 
   have $p$ recommended as a supplement for $o$ --- or as an alternative to 
   $o$. The latter would be extra useful if we have useful time estimates, 
   forthy it may e.g.\ be that $p$ fits the desired composition estimate 
   better than $o$.
   \item Authors that have module $o$ may be recommended $p$ irrespective of 
   the other modules in their composition, e.g.\ in the aforementioned time 
   constraint  scenario.
\end{itemize*}

As touched on above, combining chronology insights with other metadata, we may 
make several observations.

As an example, let $n : T$, $q : U$, and $o, p : V$, where $:$ can be read 
``has-type'', i.e.\ lhs is a module, and rhs is the type of module (e.g.\ news 
article, scientific paper, video, game, or quiz). Then let $t$ be a function 
that takes a module as its input, and returns the module's type metadata as 
its output. If a user has $n$ and $q$, we can recommend $o$ and $p$ as above. 
But $t o = t p$ means that $V$ is potentially interesting. Thus we can safely 
recommend the set of other modules with the same type as $o$ and $p$, i.e. 
$\left\{ {t m = V \vert m \in M} \right\}$, where $M$ is the set of all 
modules.

If we let $p : W$, and arrive at the conclusion that $p$ should be suggested 
from another metric --- we have that the successor of $n = o \vert p$ in the 
imagined popular canvas, so $o$ and $p$ are similar in other ways than $t o$ 
and $t p$ --- we could recommend $\left\{ t m = t o\text{, or }t m = t p \vert 
m \in M \right\}$, i.e.\ the set of all modules with either $V$ or $W$ as 
their type.

These insights hold for other things than types. As an example take $:$ to be 
``has-topic'', and $t$ to be a function from a module to its topic. The system 
could also look at a combination of different metadata to work out heuristics 
for how to suggest modules.

Satisfied with a sufficient motivation for chronology-awareness, we now turn 
to the implementation of it. The core idea of plaimi's tempuhs system seems to 
be tailored to our use case. The cons are that it needs extended 
expressiveness for the relationships of chronology elements (called timespans 
in tempuhs), and that it has not been put to the test for production use when 
several users are considered. The pros are that it is known to deal with a lot 
of data, and that it offers good guarantees of representing our data 
logically, and preserving said logic. Using tempuhs would allow us to distil 
canvases into timespans, which lets us consider the time aspect carefully. It 
is reasonable to think that several new insights are attainable if we go this 
route.

\subsection{Estimation}
\label{estimation}
In Section~\ref{chronology}, modules are considered in relation to each other. 
Thus everything is happy days. However, in this section we consider estimating 
the time a module takes, which proves to be a quite complicated endeavour.

Being able to estimate the time a module takes to complete is useful for the 
author of the module and the module users both. This would be a conservative 
augmentation both in terms of technical and philosophical impact, making it 
perhaps \emph{less interesting} than other prospects explored in this paper, 
but at the same time perhaps all the more immediately useful.

It is easy to conceive of the practical aspects of this idea. There are two 
levels to it. First, let authors of modules estimate the amount of time a 
module will take, and store it as module metadata. The design changes involved 
are quite small, the programming required is minuscule.

There is some benefit to this, but the obvious issue is that the estimator 
might be wrong. It may be wrong in several ways for several reasons. 
Conceivably the estimation reflects what the author is aiming for rather than 
what the author has actually achieved. To put it simply: they might be wrong.

The next logical step then becomes to accumulate how much time users 
\emph{actually} spend. This is more complicated to implement, but not 
\emph{too} complicated. We will require client-side executable code to achieve 
this, which means that any prevention of such code will prevent us from 
gathering useful data. This is not too worrying as there will be little reason 
for users to prevent this code from running, meaning that few users will do 
so. When viewed in isolation, the performance penalties of this code will be 
negligible.

But there are several weaknesses to the metric itself. Naïvely accumulating 
how long a user spends on a particular website results in a plethora of 
useless data. Two easily imagined extrema are users leaving a website up for a 
very long time, and users immediately leaving the website. It is similarly 
easy to imagine why this would happen. As one example of each: 1. Consider a 
primary school pupil visiting a website during the very end of instruction 
time, and leaving it up until the next instruction time (e.g.\ visiting it at 
the end of Friday, and leaving it up through the weekend). 2. Consider a user 
visiting the wrong website and immediately leaving it.

Accounting for those specific problems is non-trivial. Normally, a standard 
deviation cut-off would suffice adequately, but in our case we are faced with 
something like a tri-modal distribution. Maybe it's a leptokurtic 
distribution. That would be nice. But unfortunately, it might be, or it might 
not be. That's a lot worse. And it's about to get worse. Because it might be 
leptokurtic, and then it might change into a fat-headed distribution. And then 
it might reverse back again. And then it might become heavy-tailed. Etc. The 
modules might even have different distributions (that change as more people 
visit them.) If you have a leptokurtic distribution in two modules, a 
heavy-tailed distributed module, and two fat-headed and fat-tailed 
distributions, what then of the composition of these five modules? What then 
when the third becomes leptokurtic and the fourth becomes heavy-tailed? Doing 
this properly is not going to be trivial.

Are we having fun yet? Because it's about to get even more fun. Consider 
having a quiz about monoids in semigroup theory. The author estimates it to 
take $n$ minutes. A primary school student and a maths postgraduate go through 
the quiz. The estimation can in this situation be wrong in two ways --- too 
short and too long. This raises the question --- is the estimation in general 
useful at all?

So then the next step is to tie estimation to knowledge. I.e.\ we need to know 
how much the user knows about something, and derive estimations based on how 
much similarly knowledgeable users know about the same thing. This is 
non-trivial enough to not warrant any more examples. The observant reader 
``gets the picture'', as it were.

Let's take a step back before we end up with a horror story instead of a 
paper. Let's look at the benefits and opportunities afforded to us by 
implementing this.

Authors may receive useful analytics regarding their modules. Some banal 
analytics are ``users are spending longer on this module than you thought they 
would'', and ``these modules in your composition are approximately of the same 
length, but this fourth one takes a lot longer''. The former suffers from the 
context problem, but the latter is actually rather pleasant. Indeed most 
things we can say about one module in relation to some other modules in the 
same composition is usually immediately useful without a Ph.D in statistics.

While module authors are rewarded with feedback, module users are provided 
with useful information on how they want to spend their time, which makes our 
system interrupt flow more seldom. Users may search for modules and 
compositions based on time estimates. The same benefit applies to indirect 
users. E.g.\ classroom teachers might search for the compositions that yields 
the highest ALT to allocated time ratio.

Module authors can be said to be module users as well, in that they will often 
remix modules, and they benefit in a similar manner to module users. They can 
e.g.\ search for modules that fit their estimated composition length.

To make estimates useful for users (including indirect users and remixers) the 
programmer needs a statistics Ph.D or so to deal with transforming the data, 
and then contextualising it. To make estimates useful for authors, they will 
likely need a Ph.D themselves. We should however try our best to help them 
make sense of it. Perhaps authors should have to complete an introduction to 
statistics composition before being able to access their analytics.

\subsection{Repetition}
\label{repetition}
The law of recency states that learning degrades over time, and the law of 
exercise says that learning is increased through 
repetition\cite{murphy2011games}. A common solution to this problem is 
discussed in the paper that initially proposed the 
canvas\cite{berntsen2015enabling}, namely spaced repetition (combined with 
testing); i.e.\ studying across several separated sessions in time rather than 
spending the same amount of time in a single session.

This often leads to higher retention, and is one of the most reliable findings 
in human learning research (echoed in hundreds of studies, the first of which 
dating to the 1800s). It has been predictably demonstrated in both children 
and adults, and for both trivial knowledge (simple facts) as well as advanced 
concepts\cite{carpenter2012using}. It has also been shown to be beneficial in 
realistic (applied) contexts\cite{sobel2011spacing, carpenter2012using}.

As we've already discussed\cite{berntsen2015enabling}, spaced repetition with 
testing is a credible learning method, and thus appealing feature to include 
in our canvas. The initial canvas system was nevertheless designed without an 
emphasis on spaced repetition, in order to provide a more general framework. 
Spaced repetition is usually provided for rather specific and well-defined 
knowledge (translate this word to German, solve this equation for x, etc.), 
and arguably makes less sense for a news article leading up to a discussion. 
The canvas is merely a way to glue things together, where ``things'' is an 
ever so broad term. Another issues is that finding the optimal spacing gap is 
notoriously difficult, and inherently contextual (there is no 
``one-size-fits-all'' solution)\cite{carpenter2012using}.

As it stands, centering the entirety of the canvas's design on spaced 
repetition is unlikely. But it is altogether conceivable to augment it with a 
separate system specialising in spaced repetition. As an example, there is 
nothing that precludes the canvas from facilitating a system which focusses on 
spaced repetition.

The modules and compositions thereof would need metadata tailored to the 
spaced repetition model of learning, but this is not in itself a difficult 
task. The amount of work to at the very least be natively 
spaced-repetition-aware is very low, and the benefits may be 
disproportionately high for systems that may want to use the canvas system. 
Consequently, it is likely a good idea to make at least that much effort.

The next level of effort would be to include a way for learners to self-assess 
retention, in a manner similar to what
Anki\footnote{\url{https://ankiweb.net/about}} and similar programs do. There 
must be a way to mark a module as spaced-repetition-aware, which will then let 
the user self-assess its learning effect at the end of it. This is a feature 
which is useful regardless of spaced repetition, as we are now able to say 
something about our users' retention level based on self-assessment. By 
extension we can say something about the retention success of modules. It 
trivially also follows that we may say something about assessment, both from 
the perspective of a user, and of a module. Assessment can be made into an 
interactive and engaging affair through our system's avatar 
feature\cite{berntsen2015enabling}.

The final step is to actually encourage repeating somehow. This is likely out 
of scope for our system for now. There is however, as mentioned, nothing 
precluding an external system from augmenting our system with this.

\subsection{Synchronisation}
\label{synchronisation}
The canvas system is part of a project dubbed ``Learning by Teaching'', and 
was originally conceived as a stepping stone towards cultivating learning by 
teaching, which fosters advantages that do not manifest if the learner relies 
exclusively on an external teacher in a situated learning 
environment\cite{cortese2005learning}. Additionally, the canvas was designed 
to offer an intuitive way of graphically composing e-learning 
modules\cite{berntsen2015enabling}.

The stepping stone satiated by the canvas is authoring. As such, the canvas 
can be said to achieve ``learning by authoring'', a process that covers 
gathering of learning material, and organising. As a result, the canvas 
software focusses on authors.

However, the compositions that are made on the canvas are only interesting 
insofar as they are used. We must therefore not neglect the end-users of 
e-learning in favour of the authors. Although we wish to encourage a learning 
effect from authoring modules, there will be some pure end-users that do not 
author anything. When focussing our attention on these end-users, we must 
consider collaborative learning, as discussion is shown to foster the 
development of critical thinking\cite{gokhale1995collaborative}.

In the interest of collaboration, mechanisms for synchronising users are 
desirable. Three suggestions are discussed here;
\begin{itemize*}
  \item realtime (synchronous) collaboration,
  \item wait-for-me collaboration,
  \item and timeslot collaboration.
\end{itemize*}

Exploring realtime collaboration offers several specific features. The 
collaboration may take place on two levels --- using the modules, discussing 
the modules, or both.

First, let's talk about collaboratively using and discussing modules in 
realtime. The immediate idea here is akin to Twitch Plays Pokémon, where over 
a million users for over two weeks voted on what to do at every step of the 
game \begin{CJK}{UTF8}{min}ポケットモンスター 赤\end{CJK} (Poketto Monsut\={a} 
Aka, known as Pokémon Red outside of Japan), in a strictly egalitarian manner, 
whilst discussing the game in a live chat\cite{tpp}. The experiment is a 
significant phenomenon that demonstrates that social groups are able to unite 
in social contexts where obstacles are presented\cite{margeltwitch}. The 
experiment is largely transferable to users of our canvas system, in that 
there are several modules for which it is possible to have a group of people 
using them at the same time with e.g.\ the majority vote deciding how they 
progress. More sophisticated voting mechanisms such as Condorcet may be 
desirable to provide better overruling heuristics. Discussion may be directly 
transfered; i.e.\ a regular realtime chat is provided.

There are several ways of making this idea more sophisticated. Users may need 
to discuss and argue their views as to why e.g.\ one answer in a quiz is 
correct and others are not, in order to achieve a satisfactory outcome (per 
some voting heuristic), lest they be prevented from progressing. Discussion 
may be augmented with features that make it easy to refer to information 
within a composition. As an example, in a composition where the users are on 
module three, a quiz, they may wish to refer to module two, an article, to 
strengthen their argument. In this example the user needs a simple way of 
accessing previous modules, and a way of easily using them in a discussion.

If the reader is concerned that the idea has become \emph{too} sophisticated 
now, fear not; there are equally many ways of distilling it down into simpler 
components. E.g.\ A chat by itself. This modest feature would be a rather 
large extension of the canvas system. Especially as it was argued against in 
the original implementation to avoid abusive 
behaviour\cite{berntsen2015enabling}.

Instead of each participant actively influencing module outcomes, a seat mode 
may be used. There are several conceivable implementations of this. One is 
that there is one (somehow elected) person in control all the time, that needs 
to act on behalf of the group. Another is a hot seat solution in which the 
seat holder changes based on some heuristic.

Modules that are merely articles or videos or other non-interactive learning 
material arguably benefit the least from realtime collaboration --- fast 
readers must wait for slow readers, and that's about it. This is where 
wait-for-me collaboration becomes useful. The general idea is that there are 
several synchronisation points where users must become synchronised. In the 
example above, it would be natural that if module one and two were reading 
material, these may be pursued independently. There is nothing precluding the 
joint existence of wait-for-me and realtime mechanics, so that once the users 
are synchronised, they may use a module --- such as the quiz in module three 
--- in collaborative realtime. Another useful combination is wait-for-me 
synchronisation at certain intervals, after which realtime discussion takes 
place. But wait-for-me mechanics have useful properties when viewed 
independently as well. One concrete example that is easy to imagine useful is 
in a largely situated learning environment where it is desirable that all 
learners possess roughly the same information.

Timeslot collaboration is another useful idea for situated learning 
environments. It is additionally also useful for learners that want to 
collaborate across timezones, or following some self-imposed schedule. A 
timeslot mechanism would entail completing modules in certain timeslots. 
Teachers often set learning material per class per week in school, and gives 
homework based on rather tight timeslots, so this is a familiar concept. 
Again, it may be combined freely with the other two synchronisation methods. 
It may also be nested. E.g.\ a timeslot to do a module composition wherein 
wait-for-me mechanics are used for non-interactive learning material, 
culminating in a real-time quiz and subsequent discussion.

Where naïve wait-for-me moves in the pace of the slowest participant at the 
risk of alienating the quicker participants, naïve timeslot synchronisation 
moves at a set pace and risks leaving the slower participants behind. These 
problems mean that the features may have exactly the opposite of our intended 
effect, maximising ALT, for some subset of users. Wait-for-me synchronisation 
needs to consider a method of progressing if one (or more) participants are 
slowing the group down, whilst timeslots need to consider a way of ensuring 
that participants are actually learning. Realtime in turn risks virtually all 
known problems with online social interaction\ldots

Collaborative learning is most effective with an instructor that facilitates 
learning\cite{gokhale1995collaborative}. It then follows that we should seek 
to foster learning by instruction in addition to learning by authoring. 
Marrying the two (i.e.\ users acting as instructors of material they have 
themselves authored) gets us much closer to learning by teaching proper.

Consequently synchronisation should be extended to encompass instructors as 
well. There are several ways of achieving this. Instructors may provide 
realtime feedback whilst a group is going through a module, or discussing it. 
They may also act as the seat holder, thereby offering a potential solution to 
any social problems.

This entire section has a certain latent conjecture hanging over it: 
Synchronous collaboration is mostly interesting in a (semi) situated learning 
environment. However, this environment needn't be a classroom setting. It just
needs to be facilitated in the system that surrounds the canvas. Study groups 
or a similar mechanism in which learners may organise may be added, including 
potentially a matchmaking system, and a forum for getting in touch with 
potential collaborators. All of which are major undertakings and nigh-complete 
transformations of the original concept --- this isn't to be taken lightly.

Another problem with uncoordinated collaborators is the possibility of 
upsetting the precarious ALT by effectively making each collaborator adapt to 
each other. This could potentially result in every single collaborator 
following a sub-optimal pace, thereby harming the chances of achieving ALT, 
making this a negative feature rather than positive. Well put-together study 
groups alleviate this slightly, but not completely.

We elect not to explore potential implementation problems in detail in this 
section. Just like the core repetition idea discussed in 
Section~\ref{repetition}, the ideas presented here are of such a magnitude as 
to warrant completely new user experience research. However, it is noteworthy 
that nothing discussed in this section is fundamentally difficult from a 
technological perspective. The user-interface and -experience design 
challenges are far greater (though not insurmountable).

Like we concluded with spaced repetition, it is entirely plausible that 
synchronous collaboration is best left to another system which augments the 
canvas. It may also be suitable as a special part of some expanded system, 
wherein the learning material itself is optimised for a collaborative premise. 
Collaborative material which enables critical thinking and discussion are 
likely to be more successful than users attempting to collaborate on material 
not designed with collaboration in mind.

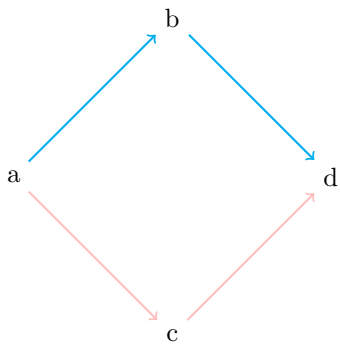
\begin{figure}[H]
\begin{centering}
\begin{tikzpicture}[node distance = 6 em, auto]
\node(a){a};
\node[right of = a](i){};
\node[above of = i](b){b};
\node[below of = i](c){c};
\node[right of = i](d){d};

\draw[cyan,thick,->](a) -- (b);
\draw[pink,thick,->](a) -- (c);
\draw[cyan,thick,->](b) -- (d);
\draw[pink,thick,->](c) -- (d);
\end{tikzpicture}
\caption{A canvas optimised for collaboration}
\label{collabcanvas}
\end{centering}
\end{figure}

With learning material optimised for collaboration, it is possible to 
approximate real world tasks to a higher degree. As an example, consider the 
software engineering composition visualised in Figure~\ref{collabcanvas}. Let 
$A$ and $B$ be the participants of this canvas. Let $a$, $b$, $c$, and $d$, be 
modules. Let $a \to b$ and $b \to d$ be flows unique to $A$, and $a \to c$ and 
$c \to d$ unique to $B$. The topic might be compilers. $a$ might be an 
introduction to compilers, then $b$ can be an introduction to frontend 
(lexing, parsing, etc.), and $c$ an introduction to backend (assembling, 
code-generation, etc.), and finally $d$ can be a quiz about both front- and 
backend fundamentals. This models real world collaboration in a sense. It is 
not unusual to divide up tasks like this in software engineering. The quiz 
might now be a realtime collaborative task in which the $A$ and $B$ must rely 
on each others' knowledge in order to pass it. Through this process, it is 
plausible that $A$ will learn about compiler backends, and that conversively 
$B$ will learn about compiler frontends.

The interactive nature of our canvas makes collaborative learning a natural 
fit. But the original design did not consider collaborative learning, and as 
such the augmentation must be considered too great to be done recklessly. User 
experience research is thus thoroughly recommended, and indeed necessary.

\section{Further research}
\label{further}
There are ideas worth exploring related to ALT that emphasise interactivity 
and assessment. The law of exercise emphasises that in order to achieve the 
best learning results, practice and feedback must 
coexist\cite{murphy2011games}.

An integral part to ALT is that the learner experiences high levels of 
success\cite{cotton1990educational, murphy2011games}. Facilitating ALT is in 
principle straight forward, but for the balancing of difficulty of skill. 
Success requires a delicate balance in which tasks are challenging yet 
achievable. Feedback (the manner in which the learner perceives their 
progress) is intrinsically entangled with achieving this balance, and 
assessment is in turn intrinsically entangled with 
feedback\cite{murphy2011games}.

It would be worthwhile further investigating augmenting non-linearity as a 
means to achieve the difficulty balance. There are numerous angles to 
investigate. For instance, modules may be interactively rearranged or 
hot-swapped based on difficulty.

In Section~\ref{repetition}, naïve self-assessment is suggested. More 
sophisticated methods of assessment are worthy of investigation. One novel 
approach to interactively tutoring learners is Ask-Elle, a programming tutor 
for the Haskell programming language which provide students with feedback on 
incomplete programs, and give hints on how to proceed in order to solve a 
programming exercise\cite{jeuring2012ask}. The avatars of our extended system 
help the system to achieve a more human touch\cite{berntsen2015enabling}, and 
are practical candidates for such a tutoring system, which might double up as 
an assessment tool.

In Section~\ref{synchronisation}, instructor-integration is briefly mentioned. 
Presently, module authors are primarily involved pre-learning. Attempts at 
involving them \emph{during} learning in an instructor role would be 
worthwhile.

\section{Conclusion}
Time as a concept is intrinsic to the canvas system proposed by plaimi, as we 
want to maximise ALT. There are several promising insights available to us by 
viewing compositions as chronologies. This simple exercise in perspective 
offers insights that particularly manifest as suggestions we may offer the 
users based on chronology metadata. Knowing which modules tend to follow 
others is simply extremely useful. It is also rather conceptually trivial. It 
does however not directly improve the ALT capabilities of our system per se.

Offering estimation mechanisms in module metadata may alleviate some 
time-management burdening for our users. It makes it easier for authors to 
find modules to fit their composition --- particularly if they are making the 
composition for a situated learning environment in which allocated time must 
be carefully considered --- and it makes it easier for module end-users to 
find suitable learning material. Allowing authors to embed man-made 
estimations as module metadata is a very modest but good extension, but the 
estimations are educated guesses at best. Furthermore, learning material time 
estimations depend heavily on the end-user. Therefore an even better extension 
would be to gather data and do contextual estimation for each user. This is 
however a very difficult problem.

Spaced repetition is accepted as often leading to higher retention and thus 
better learning. We could encode spaced repetition metadata in modules and 
compositions thereof, making our system spaced-repetition-aware, thereby 
further extending its usefulness and area of application. We could also 
implement an insight afforded from spaced repetition software --- the notion 
of self-assessment immediately post-learning. This lets us say useful things 
related to retainment and assessment regardless of spaced repetition.

Collaborative learning can offer a positive learning effect, but if not done 
properly this might be antithetical to ALT. Synchronisation for collaboration 
may be done in realtime, in wait-for-me time, or by timeslots. Realtime 
collaboration in interactive learning material is an interesting prospect. So 
is synchronising users at given intervals, especially when combined with a 
realtime module, e.g.\ evaluation (a quiz or similar) after synchronising the 
users. Timeslots may be a useful way for especially teachers in classroom 
settings to ensure that learners have similar progress. These ideas all 
present difficult problems due to their invasive nature. Further investigation 
is encouraged to take place in a separate system with learning material 
optimised for collaboration, which might to some extent marry the advantages 
of ALT and collaborative learning.

To sum up, the following features should be implemented:
\begin{itemize*}
  \item order-awareness for chronology insights,
  \item author estimation of module length,
  \item a metadata framework for spaced repetition,
  \item and post-module self-assessment capabilities.
\end{itemize*}

These are, not coincidentally, the most modest features proposed in the paper.

The following more invasive changes were discussed:

\begin{itemize*}
  \item system estimation of module length,
  \item contextualised (user-customised) estimation of module length,
  \item spaced-repetition encouraging,
  \item collaborative realtime module use,
  \item easy referring to modules and specific elements therein,
  \item wait-for-me collaboration in which users are periodically synchronised 
  to the same module,
  \item timeslot collaboration to ensure that users are somewhat synchronised,
  \item various combinations and nesting of the proposed synchronisation 
  methods,
  \item instructor-integration mechanics for collaborative learning,
  \item and a sub-system optimised for collaborative learning.
\end{itemize*}

Further research is encouraged for all of these more invasive features. 
Collaborative learning is particularly interesting due to the weight of its 
potential augmentation. It is also particularly difficult due to its potential 
negative impact on ALT.

\end{multicols}
\bibliography{paper}
\bibliographystyle{unsrt}
\newpage
\end{document}